\documentclass[%
 reprint,
 amsmath,amssymb,
 aps,
]{revtex4-2}

\usepackage{graphicx}
\usepackage{dcolumn}
\usepackage{bm}
\usepackage{hyperref}
\usepackage{slashed}
\usepackage{subfloat}
\usepackage{pgfplots}
\usepackage{subfigure}
\usepackage{etoolbox}   
\usepackage{orcidlink}
\usepackage{geometry}
\definecolor{acsblue}{RGB}{17,76,139}

\begin{document}

\fontsize{8}{9}\selectfont
\preprint{APS/123-QED}

\title{Phase-space structure and nonlinear dynamics of a charged particle on a helicoidal manifold under a magnetic field}

 
\author{Abdullah Guvendi\orcidlink{0000-0003-0564-9899}}
\email{abdullah.guvendi@erzurum.edu.tr (Corresponding Author)  }
\affiliation{Department of Basic Sciences, Erzurum Technical University, 25050, Erzurum, Türkiye}

\author{Hassan Hassanabadi\orcidlink{0000-0001-7487-6898}}
\email{hha1349@gmail.com }
\affiliation{Physics Department, California State University, Fresno, CA 93740, USA}
\affiliation{Department of Physics, Faculty of Science, University of Hradec Králové, Rokitanského 62, 500 03 Hradec Králové, Czechia}

\author{Semra Gurtas Dogan\orcidlink{0000-0001-7345-3287}}
\email{semragurtasdogan@hakkari.edu.tr}
\affiliation{Department of Medical Imaging Techniques, Hakkari University, 30000, Hakkari, Türkiye}

\author{Omar Mustafa\orcidlink{0000-0001-6664-3859}}
\email{omar.mustafa@emu.edu.tr}
\affiliation{Department of Physics, Eastern Mediterranean University, 99628, G. Magusa, north Cyprus, Mersin 10 - Türkiye}

\date{\today}

\begin{abstract}
{\fontsize{8}{9}\selectfont \setlength{\parindent}{0pt}
We analyze the classical dynamics of a charged particle constrained to a helicoidally embedded Riemannian manifold in $\mathbb{R}^3$ under a uniform magnetic field in the ambient space. The induced metric $ds^2=du^2+(1+w^2u^2)dv^2$ and the pulled-back symmetric gauge yield an exact reduction to a one-dimensional nonlinear Hamiltonian system. The resulting effective potential couples geometry and magnetic field, producing transitions between bounded and unbounded motion and a reorganization of phase-space topology. In the asymptotic regime, the dynamics reduces to a harmonic oscillator with $\omega_{\mathrm{eff}}=\omega_c/2$ and $\ell=\sqrt{2}\,\ell_\mathcal{B}$. The system admits a Landau-type semiclassical spectrum and exhibits a geometry--magnetic control parameter $\Lambda=q\mathcal{B}+\hbar k_v w$ governing a chirality transition.}
 \end{abstract}


\maketitle

\tableofcontents

\section{Introduction}

\vspace{0.01cm}
\setlength{\parindent}{0pt}

The relationship between geometry and electromagnetic dynamics in constrained systems provides a fundamental route to understanding how curvature shapes physical motion beyond flat Euclidean settings. When a charged particle is restricted to move on a curved manifold, the metric is no longer a passive background element; instead, it directly modifies the kinetic structure, reshapes phase-space evolution, and generates effective forces that resemble externally imposed potentials \cite{kolovs2023charged,narzilloev2021dynamics,aliev2002motion}. In this setting, curvature functions as a dynamical ingredient that influences confinement, transport behavior, and spectral structure in a manner that can be traced directly to the geometry of the embedding space \cite{turimov2022circular}. This perspective is particularly relevant for low-dimensional systems where geometry can be engineered to control particle motion without modifying material composition \cite{lopes2012continuum,gonzalez2010graphene,errehymy2025frame,garcia2020graphene,guvendi2023fermion,dogan2025geometric,dogan2025ray,gurtas2025ray}.

\vspace{0.01cm}

Helicoidal surfaces provide a natural and physically transparent example of geometry-controlled dynamics. These manifolds are generated by a continuous twisting of a planar strip and are characterized by a tunable geometric parameter that measures the rate of rotation along the longitudinal direction \cite{saxena-1,guvendi2025damped,guvendi2025photonic,gurtas2025ray2}. Unlike flat geometries, the induced metric on a helicoid depends explicitly on the transverse coordinate, which produces a position-dependent modification of longitudinal motion. This feature leads to a coupling between transverse and longitudinal degrees of freedom through the metric itself. Even in the absence of external fields, this structure generates nontrivial trajectories, modifies effective inertia, and produces localization effects that originate purely from the embedding geometry \cite{guvendi2025photonic,gurtas2025ray2}.

\vspace{0.01cm}

When a magnetic field is introduced, the dynamics becomes significantly richer due to the coupling between gauge structure and curved geometry \cite{background,gurtas2025twist}. In a curved space, the vector potential acquires a coordinate dependent representation once it is projected onto the tangent space of the manifold. As a result, the magnetic interaction cannot be separated from the metric structure, since both contributions are shaped by the same embedding. In helicoidal geometries, the twist parameter simultaneously controls the strength of metric deformation and the magnitude of the induced gauge potential \cite{phong2022boundary,zhang2014strain}. This leads to a combined mechanism where curvature modifies inertial response while the magnetic field shifts canonical momentum in a geometry dependent manner, producing a unified framework for confinement and transport.

\vspace{0.01cm}

Most existing studies of particle dynamics in curved geometries rely on perturbative treatments, numerical simulations, or effective approximations that simplify the underlying structure \cite{guvendi2026charged,ferrari2008schrodinger}. While these approaches provide useful insights, they often obscure the exact role of geometry in shaping phase-space organization and transition between dynamical regimes. A fully analytical treatment that retains the complete nonlinear dependence on both curvature and magnetic field allows a clearer understanding of how bounded motion, separatrix trajectories, and transport regimes emerge from the structure of the effective potential. In particular, an exact reduction of the dynamics to a single effective degree of freedom reveals how confinement is generated directly from the geometry rather than imposed through external constraints (see also \cite{guvendi2026charged}).

\vspace{0.01cm}

In this work, we develop an analytical description of the motion of a charged classical particle constrained to a helicoidally embedded Riemannian manifold under a uniform magnetic field aligned with the symmetry axis. Starting from an explicit embedding, we derive the induced metric and construct the surface projected vector potential in a consistent manner. The resulting formulation reduces the full dynamics to a single nonlinear equation governing the transverse coordinate, from which all physical regimes can be classified. We obtain exact expressions for turning points, identify conditions for bounded and unbounded motion, and demonstrate how curvature and magnetic field jointly determine the structure of phase space. In the asymptotic limit, the system reduces to a renormalized harmonic oscillator, leading to modified cyclotron dynamics and geometry dependent scaling of the magnetic length, thereby establishing a direct connection between curvature induced structure and Landau type quantization. The paper is organized as follows. Section \ref{TGNR} presents the construction of the helicoidal manifold and derives the corresponding induced metric from the embedding in $\mathbb{R}^3$. Section \ref{MF} introduces a uniform magnetic field in the ambient space and its consistent projection onto the helicoidal surface, leading to the intrinsic gauge potential. Section \ref{CHF} develops the classical Hamiltonian formulation of a charged particle on the manifold and derives the reduced effective potential governing the transverse dynamics. Section \ref{PS} analyzes the exact orbital structure, confinement conditions, and phase-space properties of the system, including turning-point analysis and dynamical regimes. Section \ref{quantization} discusses semiclassical quantization, asymptotic spectral behavior, and the emergence of a geometry-induced quantum phase transition. Finally, Section \ref{conc} summarizes the main results and outlines possible extensions to other geometrically constrained and gauge-coupled systems.

\section{Geometric structure of the helicoidal manifold}\label{TGNR}

\vspace{0.01cm}
\setlength{\parindent}{0pt}

We construct the electromagnetic field on the helicoidal manifold by solving Maxwell’s equations in the ambient Euclidean space $\mathbb{R}^3$ \cite{background,gurtas2025twist} and subsequently pulling back the resulting vector potential to the embedded surface via the immersion $\mathcal{X}(u,v)$. This ensures consistency with the induced metric structure of the helicoidal manifold. The ambient space is the flat Euclidean manifold $(\mathbb{R}^3,\delta)$ equipped with global Cartesian coordinates $(x,y,z)$ and line element \(ds^2 = dx^2 + dy^2 + dz^2\). This Euclidean metric provides the reference geometry of the ambient space in which the manifold is embedded, and all intrinsic properties, including anisotropy and intrinsic curvature, arise exclusively from the embedding rather than from any externally imposed background structure. The surface is realized as a smooth immersion $\mathcal{X}:\mathbb{R}^2 \to \mathbb{R}^3$, parameterized by intrinsic coordinates $(u,v)$, where $u$ denotes the transverse coordinate and $v$ denotes the longitudinal coordinate along the symmetry axis. The embedding is defined by the helicoidal map \cite{background,gurtas2025twist}
\begin{equation}
x(u,v)=v,\quad y(u,v)=u\cos(wv),\quad z(u,v)=u\sin(wv),
\label{eq:embedding}
\end{equation}
where the constant twist parameter is defined as
\begin{equation}
w=\frac{2\pi m}{L}.\label{twist-density}
\end{equation}
Here $m$ denotes the total number of full $2\pi$ rotations accumulated along the longitudinal extent $L$, while $w$ is a geometric twist density with dimensions of inverse length. The tangent structure is obtained by differentiating the embedding map. The transverse tangent vector is
\begin{equation}
\partial_u \vec r = (0,\cos(wv),\sin(wv)),
\end{equation}
and the longitudinal tangent vector is
\begin{equation}
\partial_v \vec r = (1,-uw\sin(wv),uw\cos(wv)).
\end{equation}
The induced metric is defined as
\begin{equation}
g_{ab}=\partial_a \vec r \cdot \partial_b \vec r,\qquad a,b\in\{u,v\}.
\end{equation}
The metric components are
\begin{equation}
g_{uu}=1,\qquad g_{uv}=0,\qquad g_{vv}=1+w^2u^2.
\end{equation}
Hence the induced line element becomes \cite{background,gurtas2025twist}
\begin{equation}
ds^2 = du^2 + \chi(u)\,dv^2,\qquad \chi(u)=1+w^2u^2.
\label{eq:metric}
\end{equation}
\begin{figure}[ht]
\centering
\includegraphics[scale=0.50]{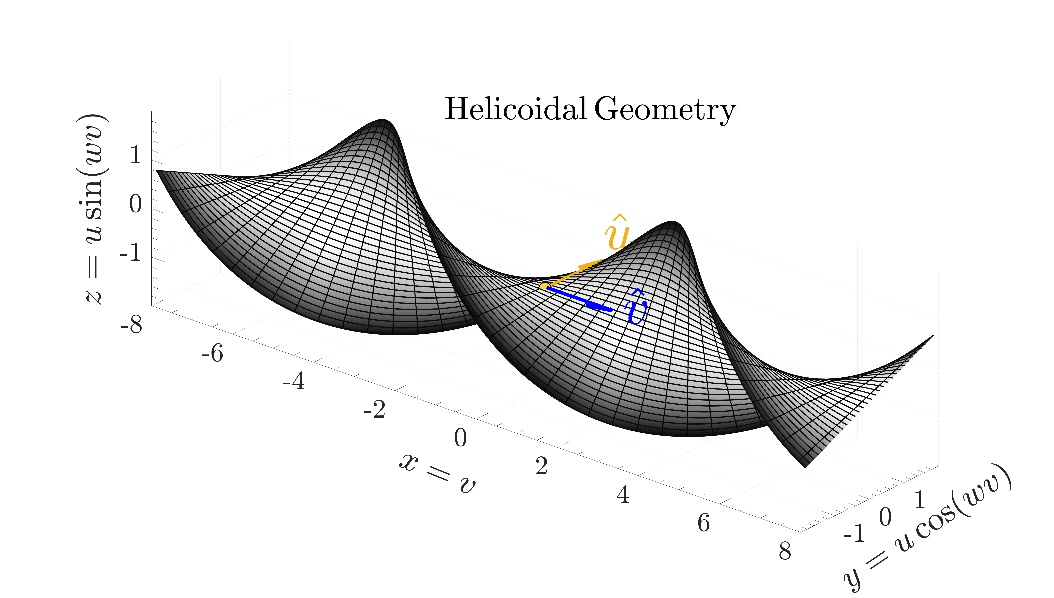}
\caption{\fontsize{7.6}{8.6}\selectfont
Intrinsic geometry of a helicoidal Riemannian manifold embedded in $\mathbb{R}^3$. The embedding is defined in \eqref{eq:embedding}, where the transverse coordinate satisfies $u\in[-\mathcal{D}/2,\mathcal{D}/2]$ with width $\mathcal{D}=2$, and the longitudinal coordinate spans $v\in[-6,6]$. The structure is characterized by a twist density $w=2\pi m/L \approx 0.503$. The induced metric is $ds^2 = du^2 + \chi(u)\,dv^2$, where $\chi(u)=1+w^2u^2$ encodes the position-dependent stretching of longitudinal line elements due to the helicoidal embedding. The basis vectors $\hat{u}$ and $\hat{v}$ denote the transverse and longitudinal tangent directions.}
\label{fig:3D}
\end{figure}
The function $\chi(u)$ acts as a position-dependent metric factor controlling longitudinal stretching. Although the metric is diagonal, the dependence of $g_{vv}$ on $u$ generates nontrivial Christoffel symbols, leading to coupling between transverse and longitudinal dynamics in geodesic motion. Consequently, motion on the surface is intrinsically anisotropic. The parameters $(m,L)$ determine the twist via \eqref{twist-density}, while $u\in[-\mathcal{D}/2,\mathcal{D}/2]$ defines the transverse domain. In dynamical applications, the Laplace--Beltrami operator associated with this metric governs kinetic motion, and thin-layer quantization may introduce additional curvature-induced geometric potentials depending on the embedding and confinement procedure. This construction provides a minimal and exact geometric representation of a helicoidal Riemannian manifold embedded in $\mathbb{R}^3$ (see Figure \ref{fig:3D}), forming a foundation for analyzing geometry-induced anisotropic transport.

\section{Uniform Magnetic field structure on the helicoidal manifold}\label{MF}

We construct the electromagnetic field on the helicoidal manifold in a fully consistent manner by starting from Maxwell's equations in the ambient Euclidean space $\mathbb{R}^3$ and then pulling back the resulting vector potential to the embedded surface via the immersion $\mathcal{X}(u,v)$. This procedure avoids ambiguities associated with defining the curl operator directly in intrinsic coordinates and ensures that the magnetic field remains well-defined in the embedding space. We consider a uniform magnetic field directed along the axis of the helicoidal structure,
\begin{equation}
\vec{\mathcal{B}} = \mathcal{B} \hat{x}, \qquad \mathcal{B} = \text{constant}.
\end{equation}
In the ambient Euclidean space, the electromagnetic field is defined through the vector potential $\vec{A}$ satisfying
\begin{equation}
\nabla \times \vec{\mathcal{A}} = \vec{\mathcal{B}}.
\end{equation}
A convenient symmetric gauge that exactly reproduces a uniform magnetic field along the $x$-direction is
\begin{equation}
\vec{\mathcal{A}} = \frac{\mathcal{B}}{2}(0,-z,y),\label{car-vec}
\end{equation}
which can be directly verified to satisfy $\nabla \times \vec{A} = \mathcal{B} \hat{x}$. In Cartesian components this reads
\begin{equation}
\mathcal{A}_x = 0,\qquad \mathcal{A}_y = -\frac{\mathcal{B}}{2}z,\qquad \mathcal{A}_z = \frac{\mathcal{B}}{2}y.
\end{equation}
We now restrict this vector potential to the helicoidal surface defined by the embedding \eqref{eq:embedding}. Substituting the embedding into the Cartesian vector potential yields the surface-restricted field
\begin{equation}
\mathcal{A}_x = 0,\qquad \mathcal{A}_y = -\frac{\mathcal{B}}{2}u\sin(wv),\qquad \mathcal{A}_z = \frac{\mathcal{B}}{2}u\cos(wv).
\end{equation}
The geometry of the surface is encoded in its tangent structure. The tangent vectors are obtained directly from the embedding as
\begin{equation}
\partial_u \vec{r} = (0,\cos(wv),\sin(wv)),\quad \partial_v \vec{r} = (1,-wu\sin(wv),wu\cos(wv)).
\end{equation}
From these, the induced metric components follow as
\begin{equation*}
g_{uu}=\partial_u \vec{r} \cdot \partial_u \vec{r}= 1,\quad g_{uv}=\partial_u \vec{r} \cdot \partial_v \vec{r}= 0,\quad g_{vv}=\partial_v \vec{r} \cdot \partial_v \vec{r}= 1+w^2u^2,
\end{equation*}
showing that the helicoidal geometry induces a position-dependent stretching along the longitudinal direction while preserving orthogonality between transverse and longitudinal directions. To obtain the electromagnetic gauge field as experienced by particles constrained to the surface, we project the ambient vector potential onto the tangent basis. The intrinsic components are defined by
\begin{equation}
\mathcal{A}_u = \vec{\mathcal{A}}\cdot \partial_u \vec{r},\qquad \mathcal{A}_v = \vec{\mathcal{A}}\cdot \partial_v \vec{r}.
\end{equation}
The transverse component is obtained as
\begin{align}
\mathcal{A}_u &= \mathcal{A}_y \cos(wv) + \mathcal{A}_z \sin(wv) \nonumber \\
&= \left(-\frac{\mathcal{B}}{2}u\sin(wv)\right)\cos(wv) + \left(\frac{\mathcal{B}}{2}u\cos(wv)\right)\sin(wv),
\end{align}
which vanishes identically, \(\mathcal{A}_u = 0\). The longitudinal component is computed as
\begin{align}
\mathcal{A}_v &= \mathcal{A}_y(-wu\sin(wv)) + \mathcal{A}_z(wu\cos(wv))\nonumber \\
&= \left(-\frac{\mathcal{B}}{2}u\sin(wv)\right)(-wu\sin(wv)) + \left(\frac{\mathcal{B}}{2}u\cos(wv)\right)(wu\cos(wv))\nonumber  \\
&= \frac{\mathcal{B}}{2}wu^2\left[\sin^2(wv)+\cos^2(wv)\right]\Rightarrow \mathcal{A}_v = \frac{\mathcal{B}}{2}wu^2.
\end{align}
Therefore, the intrinsic vector potential on the helicoidal manifold is
\begin{equation}
\mathcal{A}_u = 0,\qquad \mathcal{A}_v = \frac{\mathcal{B}}{2}wu^2.\label{vec-pot}
\end{equation}
It is important to emphasize that this result originates from the exact Cartesian solution of Maxwell's equations in flat space, which satisfies $\nabla \times \vec{\mathcal{A}} = \mathcal{B} \hat{x}$. The restriction to the helicoidal surface does not alter the physical magnetic field in the embedding space; it only changes its coordinate representation on the manifold.
The gauge-invariant electromagnetic field on the surface is obtained from the field strength tensor \(F_{uv} = \partial_u \mathcal{A}_v - \partial_v \mathcal{A}_u\). Since $\mathcal{A}_u = 0$ and $\mathcal{A}_v = \frac{\mathcal{B}}{2}wu^2$, one obtains
\begin{equation}
F_{uv} = \partial_u \mathcal{A}_v = \mathcal{B}\,wu.
\end{equation}
This quantity represents the coordinate field strength on the surface; the corresponding physical flux density is obtained after accounting for the induced metric measure. The physically invariant magnetic flux density on the helicoidal manifold is given by
\begin{equation}
\mathcal{B}_{\mathrm{surf}}(u) = \frac{F_{uv}}{\sqrt{g}},\qquad  \sqrt{g} = \sqrt{1 + w^2 u^2},
\end{equation}
yielding
\begin{equation}
\mathcal{B}_{\mathrm{surf}}(u)
=
\frac{\mathcal{B}wu}{\sqrt{1 + w^2 u^2}}.
\end{equation}
The physical magnetic field in the embedding space remains strictly uniform, $\vec{\mathcal{B}} = \mathcal{B}\hat{x}$. The $u$ dependence of $\mathcal{A}_v$, $F_{uv}$, and $\mathcal{B}_{\mathrm{surf}}(u)$ arises entirely from the geometry of the helicoidal embedding and the induced coordinate system on the surface, rather than from any spatial variation of the physical magnetic field. Finally, the twist parameter $w=2\pi m/L$ controls the strength of this geometry-induced coupling, and larger values of $w$ enhance the magnitude of the effective flux density $\mathcal{B}_{\mathrm{surf}}(u)$ experienced by particles constrained to the helicoidal manifold.

\section{Classical Hamiltonian formulation on the helicoidal manifold}\label{CHF}

\begin{figure}[ht]
\centering
\includegraphics[scale=0.64]{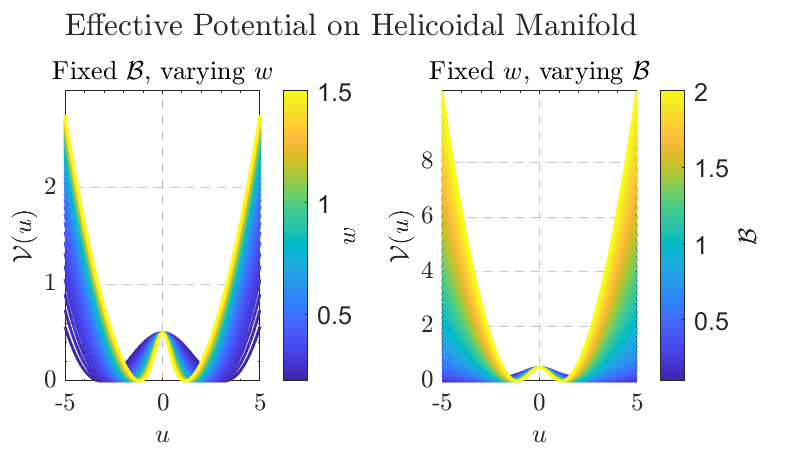}
\caption{\fontsize{7.6}{8.6}\selectfont
Effective potential $\mathcal{V}(u)$ for a classical charged particle of mass $\mu=1$ and charge $q=1$ constrained on a helicoidal manifold with induced metric. The longitudinal momentum is fixed at $\mathcal{P}_v=1$, and the dynamics are governed by the Hamiltonian in Eq.~\eqref{Hamiltonian}. The system is subjected to a uniform magnetic field $\mathcal{B}\in[0.1,2.0]$, while the intrinsic vector potential obtained from the projection of the symmetric gauge in Eq.~\eqref{car-vec} onto the helicoidal surface is given in Eq.~\eqref{vec-pot}. The transverse coordinate spans $u\in[-5,5]$, and the twist parameter $w=2\pi m/L$ is varied within $w\in[0.2,1.5]$. Left panel: effective potential profiles for fixed magnetic field $\mathcal{B}=1.0$ with varying twist parameter $w$. Right panel: effective potential profiles for fixed twist $w=0.8$ with varying magnetic field strength $\mathcal{B}$. }
\label{fig:-1}
\end{figure}
We consider a classical charged particle of mass $\mu$ and charge $q$ constrained to move on a helicoidal surface embedded in $\mathbb{R}^3$. The configuration space is the two-dimensional Riemannian manifold equipped with the induced metric obtained from the embedding in Eq. \eqref{eq:embedding}. According to the induced metric, the transverse coordinate $u$ measures intrinsic distance along the transverse direction, while the longitudinal coordinate $v$ acquires a position-dependent stretching through $\chi(u)$. The kinetic energy of the particle is fully determined by the metric and takes the form \cite{goldstein1950classical}
\begin{equation}
T=\frac{\mu}{2}\left(\dot{u}^2+\chi(u)\dot{v}^2\right).
\end{equation}
Electromagnetic interaction is introduced through minimal coupling with a surface-projected vector potential. We consider a uniform magnetic field in the ambient space directed along the helicoidal axis. Its restriction to the surface yields an intrinsic gauge structure. In a gauge adapted to the surface, one obtains \(\mathcal{A}_u=0\) so that all electromagnetic effects are encoded in $\mathcal{A}_v(u)$. Accordingly, the Lagrangian becomes \cite{goldstein1950classical}
\begin{equation}
\mathcal{L}=\frac{\mu}{2}\left(\dot{u}^2+\chi(u)\dot{v}^2\right)+q\,\mathcal{A}_v(u)\dot{v}.
\end{equation}
The canonical momenta are obtained as \cite{goldstein1950classical}
\begin{equation}
p_u=\frac{\partial \mathcal{L}}{\partial \dot{u}}=\mu\dot{u},\qquad
p_v=\frac{\partial \mathcal{L}}{\partial \dot{v}}=\mu\chi(u)\dot{v}+q\mathcal{A}_v(u).
\end{equation}
Since $v$ is a cyclic coordinate, the associated canonical momentum remains invariant under the Hamiltonian flow,
\begin{equation}
p_v=\mathcal{P}_v=\mathrm{const}.
\end{equation}
Solving for the longitudinal velocity gives
\begin{equation}
\dot{v}=\frac{\mathcal{P}_v-q\mathcal{A}_v(u)}{\mu\chi(u)}.\label{long-vel}
\end{equation}
The Hamiltonian is obtained by the Legendre transformation \cite{goldstein1950classical}
\begin{equation}
\mathcal{H}=p_u\dot{u}+p_v\dot{v}-\mathcal{L}.
\end{equation}
Substituting the velocity expressions yields
\begin{equation}
\mathcal{H}=\frac{p_u^2}{2\mu}+\frac{\left(\mathcal{P}_v-q\mathcal{A}_v(u)\right)^2}{2\mu\chi(u)}.
\end{equation}
The dynamics reduces exactly to an effective one-dimensional system in the transverse coordinate,
\begin{equation}
\mathcal{H}_{\mathrm{eff}}(u,p_u)=\frac{p_u^2}{2\mu}+\mathcal{V}(u),\label{Hamiltonian}
\end{equation}
with effective potential
\begin{equation}
\mathcal{V}(u)=\frac{\left(\mathcal{P}_v-q\mathcal{A}_v(u)\right)^2}{2\mu\chi(u)}.\label{eff-pot}
\end{equation}
Hamilton’s equations are
\begin{equation}
\dot{u}=\frac{p_u}{\mu},\qquad \dot{p}_u=-\frac{d\mathcal{V}}{du},\qquad \dot{v}=\frac{\mathcal{P}_v-q\mathcal{A}_v(u)}{\mu\chi(u)}.\label{HE}
\end{equation}
The total energy is conserved,
\begin{equation}
\mathcal{E}=\mathcal{H}=\frac{p_u^2}{2\mu}+\mathcal{V}(u).
\end{equation}
The allowed region of motion is determined by \cite{guvendi2026charged}
\begin{equation}
\mathcal{E}\geq \mathcal{V}(u),
\end{equation}
and the turning points satisfy 
\begin{equation}
\mathcal{E}=\mathcal{V}(u),\label{TP}
\end{equation}
which correspond to $p_u=0$. The dynamics is controlled by two ingredients: the induced geometry through the metric factor $\chi(u)$, and the electromagnetic interaction through the intrinsic gauge potential $\mathcal{A}_v(u)$. The twist parameter $w$ controls the strength of the geometric deformation through $\chi(u)$, while the magnetic field enters via $\mathcal{A}_v(u)$, producing a gauge-dependent shift of the conserved canonical momentum through minimal coupling. This leads to a configuration in which the transverse motion is governed by a nonlinear effective potential arising from the combined effects of geometry and electromagnetic coupling. In Figure \ref{fig:-1}, the effective potential is determined by the competition between the geometric factor $\chi(u)=1+w^2u^2$ and the magnetic contribution $\mathcal{A}_v(u)$ in \eqref{vec-pot}. Increasing the twist parameter $w$ enhances the growth rate of $\chi(u)$ with respect to $u$, which suppresses the effective kinetic contribution through the denominator while simultaneously amplifying the magnetic contribution via the $w$-dependent scaling of $\mathcal{A}_v(u)$ inside the minimal-coupling momentum shift. Increasing the magnetic field $\mathcal{B}$ enhances the magnetic contribution entering through the minimal-coupling momentum shift, leading to a steeper effective potential. The longitudinal momentum $\mathcal{P}_v$ shifts the position of the minimum of the effective potential without changing its asymptotic scaling. The resulting structure shows that geometry-induced effects governed by the twist parameter $w$ and magnetic effects governed by $\mathcal{B}$ enter the Hamiltonian through distinct functional channels, with geometry controlling the metric factor and the magnetic field controlling the effective momentum shift.

\section{Exact orbital structure, phase-space dynamics, and geometric confinement mechanisms} \label{PS}

\begin{figure*}[ht]
\centering
\includegraphics[scale=0.62]{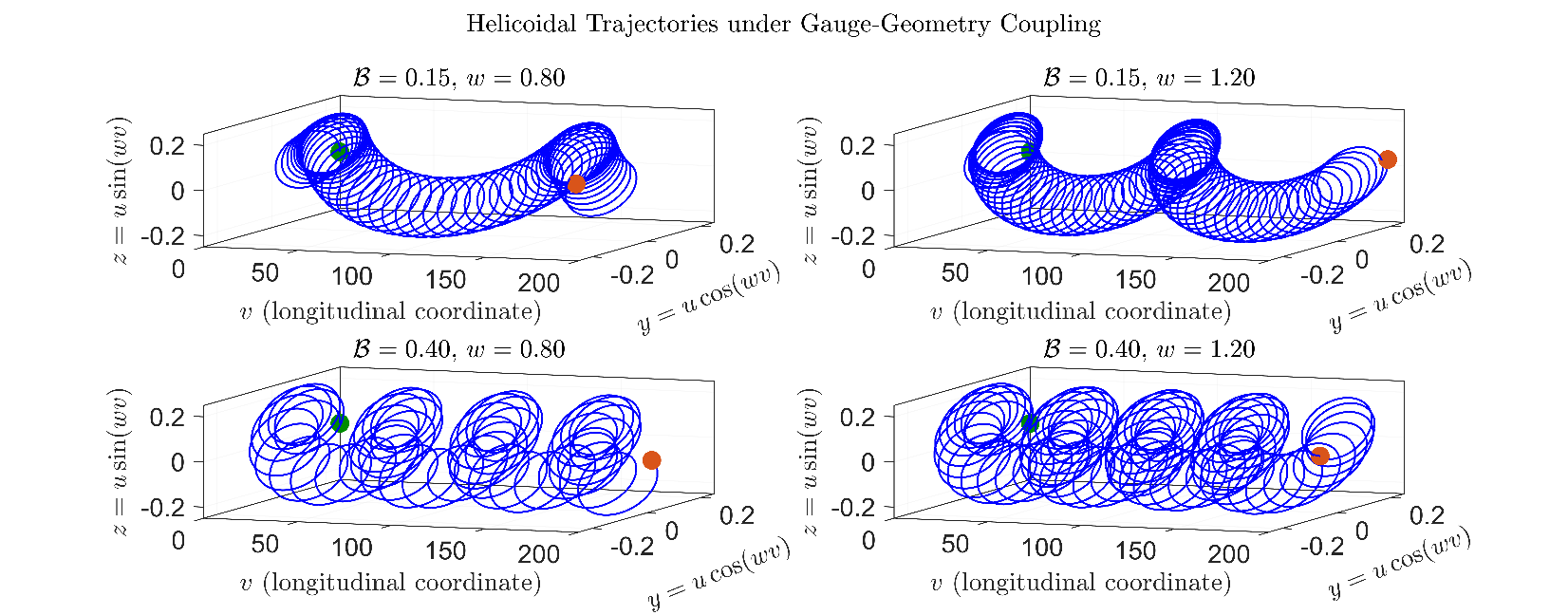}
\caption{\fontsize{7.6}{8.6}\selectfont
Trajectories of a charged particle constrained on a helicoidal manifold governed by the Hamiltonian in Eq.~\eqref{he} with vector potential in Eq.~\eqref{vec-pot}. Parameters are fixed as $\mu=1$, $q=1$, $P_v=2.0$, $\Delta t=0.001$, and $N=10^5$. The helicoidal embedding is given in Eq. \eqref{eq:embedding}. The four panels correspond to parameter sets $(\mathcal{B},w)$: (i) $(0.15,0.8)$, (ii) $(0.15,1.2)$, (iii) $(0.40,0.8)$, and (iv) $(0.40,1.2)$, showing increasing competition between magnetic confinement and geometric twisting. Initial conditions are $(u_0,p_{u0},v_0)=(0.25,0,0)$.}
\label{fig:trajectories}
\end{figure*}
We consider the motion of a charged particle constrained to a helicoidal manifold with induced metric \eqref{eq:metric} and subjected to a uniform magnetic field directed along the symmetry axis of the embedding space. The dynamics is formulated entirely in intrinsic coordinates \((u,v)\). The trajectory is described by \((u(t),v(t))\), and elimination of the explicit time parameter is achieved via the kinematic identity
\begin{equation}
\frac{dv}{du}=\frac{\dot{v}}{\dot{u}}.
\label{eq:ratio}
\end{equation}
Within the Hamiltonian formulation on the curved manifold, the canonical equations of motion take the form in Eq. \eqref{eq:metric}. This yields the reduced configuration-space flow equation
\begin{equation}
\frac{dv}{du}=
\frac{\mathcal{P}_v-q\mathcal{A}_v(u)}{\chi(u)p_u}.
\label{eq:orbit1}
\end{equation}
The gauge structure was obtained previously by projecting the symmetric gauge onto the helicoidal embedding. This form reflects the nonlinear geometric modulation of the projected gauge field induced by the helicoidal embedding. The Hamiltonian becomes \cite{goldstein1950classical}
\begin{equation}
\mathcal{H}=\frac{p_u^2}{2\mu}+
\frac{\left(\mathcal{P}_v-q\mathcal{A}_v(u)\right)^2}{2\mu\chi(u)}
=\mathcal{E},\label{he}
\end{equation}
which defines the effective potential landscape
\begin{equation}
\mathcal{V}(u)=
\frac{\left(\mathcal{P}_v-q\frac{\mathcal{B}}{2}wu^2\right)^2}{2\mu(1+w^2u^2)}.
\label{eq:Vfinal}
\end{equation}
The transverse momentum follows as
\begin{equation}
p_u(u)=\pm\sqrt{2\mu\left(\mathcal{E}-\mathcal{V}(u)\right)}.
\label{eq:pu}
\end{equation}
The reality condition \(p_u^2\ge 0\) imposes the constraint
\(
\mathcal{E}\ge \mathcal{V}(u)
\),
which defines the classically accessible configuration-space domain and fully determines confinement. Substituting Eq.~\eqref{eq:pu} into Eq.~\eqref{eq:orbit1} yields the exact reduced orbital equation \cite{goldstein1950classical}
\begin{equation}
\frac{dv}{du}=
\pm\,\frac{\mathcal{P}_v-q\frac{\mathcal{B}}{2}wu^2}
{(1+w^2u^2)\sqrt{2\mu\left(\mathcal{E}-\mathcal{V}(u)\right)}}.
\label{eq:orbit2}
\end{equation}
Integration produces the exact configuration-space orbit \cite{goldstein1950classical}
\begin{equation}
v(u)=v_0\pm\int_{u_0}^{u}
\frac{\mathcal{P}_v-q\frac{\mathcal{B}}{2}w\tilde{u}^2}
{(1+w^2\tilde{u}^2)\sqrt{2\mu\left(\mathcal{E}-\mathcal{V}(\tilde{u})\right)}}
\,d\tilde{u}.
\label{eq:orbitint}
\end{equation}
The numerical realizations of these trajectories are shown in Figure~\ref{fig:trajectories}. The structure of the trajectories reflects a nontrivial relationship between geometry-induced inertial modulation and gauge-induced momentum deformation. The metric factor \(\chi(u)=1+w^2u^2\) acts as a geometric renormalization of longitudinal inertia, effectively suppressing transport in the \(v\)-direction as transverse excursions increase. Meanwhile, the quadratic gauge structure \(\mathcal{A}_v(u)\propto u^2\) introduces a nonlinear momentum shift that increasingly dominates at large \(u\), producing strong asymmetry in phase-space flow. In weak magnetic regimes, motion is predominantly controlled by geometry, where \(w\) governs helicoidal tightness through the growth rate of \(\chi(u)\). As \(\mathcal{B}\) increases, the gauge contribution becomes comparable to or dominant over the metric term, leading to enhanced confinement and reduced radial spread. In the strongly coupled regime, nonlinear feedback between curvature and gauge coupling produces highly constrained trajectories characterized by rapid phase-space bending and suppressed transverse diffusion. Turning points are determined by the condition \(\dot{u}=0\), equivalently \(p_u=0\), yielding the exact algebraic constraint
\begin{figure}[ht]
\centering
\includegraphics[scale=0.55]{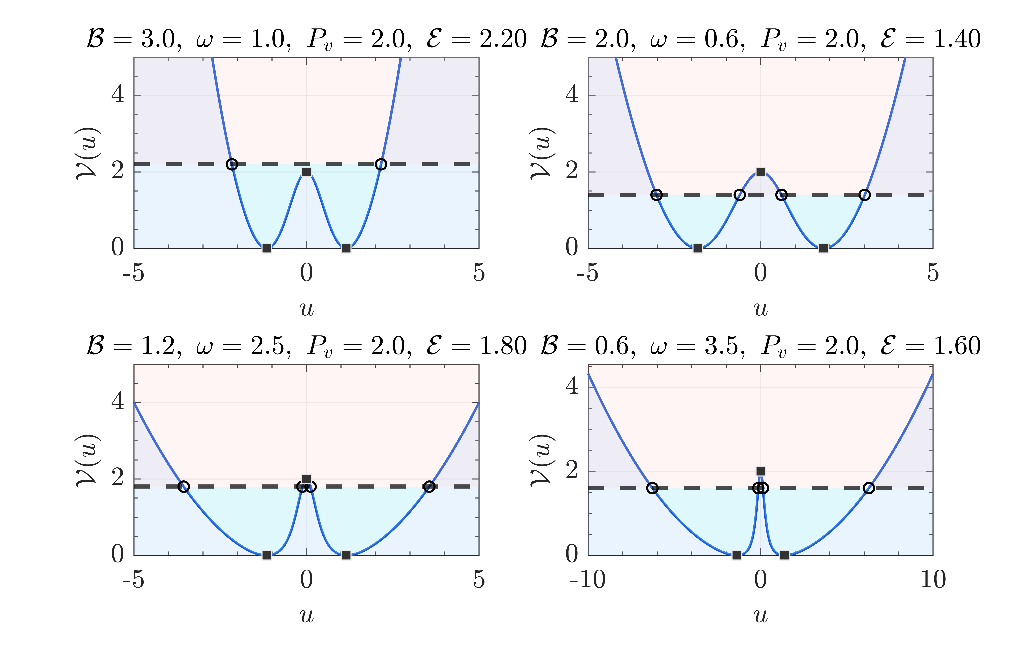}
\caption{\fontsize{7.6}{8.6}\selectfont Effective potential structure $\mathcal{V}(u)$ and confinement windows for a charged particle constrained on a helicoidal manifold governed by the Hamiltonian. The system parameters are fixed as $\mu=1$, $q=1$, $P_v=2.0$, while the energy values $\mathcal{E}$ vary across panels as specified. The geometric coupling is controlled by the twist parameter $w$ and the magnetic field strength $\mathcal{B}$. Each panel corresponds to a parameter set $(\mathcal{B}, w, \mathcal{E})$: (i) $(3.0,1.0,2.2)$, (ii) $(2.0,0.6,1.4)$, (iii) $(1.2,2.5,1.8)$, (iv) $(0.6,3.5,1.6)$. The blue curve represents the effective potential $\mathcal{V}(u)$, while the dashed horizontal line denotes the total energy $\mathcal{E}$. The shaded blue regions indicate classically allowed domains satisfying $\mathcal{E} \ge \mathcal{V}(u)$, whereas the soft red regions correspond to classically forbidden regions. Black circular markers denote turning points defined by $\mathcal{E}=\mathcal{V}(u)$, and square markers indicate stationary points satisfying $\frac{d\mathcal{V}}{du}=0$. The number of confinement windows is obtained from the connected components of the classically allowed region in configuration space.}
\label{fig:windows2}
\end{figure}

\begin{figure*}[ht]
\centering
\includegraphics[scale=0.55]{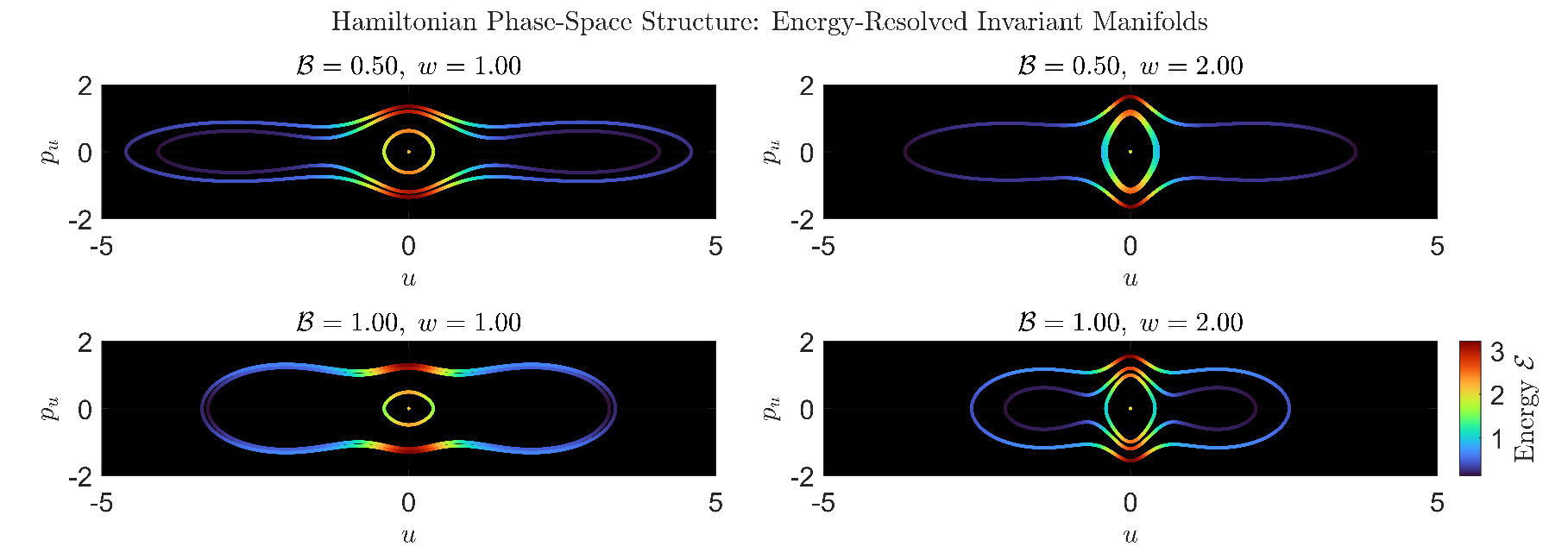}
\caption{\fontsize{7.6}{8.6}\selectfont Energy-resolved phase-space structure $(u,p_u)$ of a charged particle constrained on a helicoidal manifold under combined curvature-magnetic coupling. The dynamics is generated by the Hamiltonian in \eqref{he}, with \eqref{vec-pot}, $\mu=1$, $q=1$, and $P_v=2.0$. The phase space is computed using a symplectic velocity-Verlet integrator with time step $\Delta t=10^{-3}$ over $N=6\times10^4$ iterations. Each panel corresponds to a distinct coupling regime $(\mathcal{B},w)$: $(0.5,1)$, $(0.5,2)$, $(1,1)$, and $(1,2)$, representing increasing interplay between magnetic confinement and geometric twisting of the helicoidal embedding. Initial conditions are sampled from a uniform ensemble in $(u_0,p_{u0})\in[-0.4,0.4]\times[-1.2,1.2]$. Phase-space points are colored by the instantaneous Hamiltonian energy in \eqref{he}, revealing the foliation of phase space into energy shells. The color gradient therefore encodes the deformation of constant-energy manifolds induced by the competition between curvature-driven inertial renormalization and gauge-induced nonlinear momentum shifting.}
\label{fig:phase-space}
\end{figure*}
\begin{equation}
2\mu\mathcal{E}(1+w^2u^2)=\left(\mathcal{P}_v-q\frac{\mathcal{B}}{2}wu^2\right)^2.
\label{eq:tp}
\end{equation}
Introducing the variable \(x=u^2\ge 0\), the turning-point condition in Eq.~\eqref{eq:tp} becomes a quadratic equation in \(x\),
\begin{equation}
ax^2+bx+c=0,
\end{equation}
where
\begin{align}
a = \frac{q^2\mathcal{B}^2 w^2}{4}>0,\quad
b = -\left(q\mathcal{B}wP_v + 2\mu \mathcal{E} w^2\right)<0,\quad
c = P_v^2 - 2\mu \mathcal{E}.
\end{align}
The quadratic equation admits two algebraic roots
\begin{equation}
x_{\pm}=\frac{-b\pm\sqrt{\Delta}}{2a},\qquad \Delta=b^2-4ac.
\end{equation}
Since \(a>0\), the ordering \(x_+\ge x_-\) holds. The physical coordinate satisfies \(x=u^2\ge 0\), therefore only non-negative roots in \(x\) generate admissible turning points in configuration space, and each admissible root \(x_i\ge 0\) produces a symmetric pair \(u=\pm\sqrt{x_i}\). Using \(b<0\) gives
\begin{equation}
x_{\pm}=\frac{|b|\pm\sqrt{\Delta}}{2a},\qquad |b|=q\mathcal{B}wP_v+2\mu \mathcal{E}w^2>0.
\end{equation}
Hence \(x_+>0\) always, while \(x_-\) may be positive or negative depending on parameters. If \(x_-\ge 0\), both roots contribute and four turning points exist,
\begin{equation}
u=\pm\sqrt{x_-},\qquad u=\pm\sqrt{x_+}.
\end{equation}
The condition \(x_-\ge 0\) is equivalent to
\begin{equation}
|b|\ge \sqrt{\Delta}.
\end{equation}
Squaring and simplifying yields the exact criterion \(c\ge 0\), i.e.
\begin{equation}
P_v^2 \ge 2\mu \mathcal{E}.
\end{equation}
If \(c<0\), then \(x_-<0<x_+\) and only one root is physically admissible, giving two turning points
\begin{equation}
u=\pm\sqrt{x_+}.
\end{equation}
At the boundary \(c=0\) corresponding to \(P_v^2=2\mu \mathcal{E}\), the quadratic factorizes as
\begin{equation}
x(ax+b)=0,
\end{equation}
yielding
\begin{equation}
x_1=0,\qquad x_2=-\frac{b}{a},
\end{equation}
and the turning points become
\begin{equation}
u=0,\qquad u=\pm\sqrt{-\frac{b}{a}}.
\end{equation}
The discriminant \(\Delta=b^2-4ac\) guarantees the existence of real algebraic solutions of the quadratic equation \(ax^2+bx+c=0\) in the auxiliary variable \(x=u^2\). However, physical turning points in configuration space are further restricted by the constraint \(x\ge 0\), since \(x=u^2\) must remain non-negative. Consequently, not all algebraic roots correspond to admissible classical turning points. For \(\Delta>0\), the quadratic equation admits two distinct real roots
\[
x_{\pm}=\frac{-b\pm\sqrt{\Delta}}{2a}.
\]
Each root contributes to a physically admissible turning point only if it satisfies \(x_{\pm}\ge 0\), in which case it generates a symmetric pair \(u=\pm\sqrt{x_{\pm}}\). Depending on parameter values, this yields either two or four turning points, corresponding to one or two disjoint classically allowed regions in configuration space. For \(\Delta=0\), the quadratic equation possesses a degenerate root
\[
x_*=-\frac{b}{2a}.
\]
In this case, the effective energy curve is tangent to the effective potential in the reduced variable \(x\), and the two turning points coalesce into a single critical contact point in configuration space. If this degenerate root satisfies \(x_*\ge 0\), the corresponding trajectory lies at the boundary between qualitatively distinct phase-space behaviors, separating bounded oscillatory motion from unbounded or escaping motion. In this sense, the condition \(\Delta=0\) defines a critical separatrix trajectory in the reduced dynamics, marking the transition between disconnected invariant regions of motion. For \(\Delta<0\), no real roots exist, implying that the equation \(\mathcal{E}=\mathcal{V}(u)\) has no solution in real configuration space. As a result, there are no classical turning points, and the motion is unbounded in the transverse direction. Using the explicit coefficients of the effective potential, the root structure can be re-expressed in terms of the system parameters. The condition for admissible turning-point structure is governed jointly by \(\Delta\) and the physical constraint \(x=u^2\ge 0\). In particular, the separatrix corresponds to the condition \(\Delta=0\), where the two turning points coalesce and the energy level becomes tangent to the effective potential. This leads to the following classification:
\begin{itemize}
\item If \(P_v^2 > 2\mu \mathcal{E}\), then \(c>0\) and both roots \(x_{\pm}\) can be non-negative depending on parameter values, leading to four turning points and multiple bounded regions in configuration space.
\item If \(P_v^2 = 2\mu \mathcal{E}\), then \(c=0\) and the quadratic degenerates into \(x(ax+b)=0\), producing a boundary case with a root at \(x=0\) and an additional finite root; this corresponds to a special parameter configuration but does not, in general, coincide with the separatrix condition.
\item If \(P_v^2 < 2\mu \mathcal{E}\), then \(c<0\) and only one physically admissible root remains, yielding two turning points and a single connected classically allowed region.
\end{itemize}
Therefore, the discriminant \(\Delta\) determines the algebraic existence of candidate roots, while the physical phase-space structure is determined jointly by \(\Delta\) and the admissibility condition \(x=u^2\ge 0\). The separatrix arises from the critical condition \(\Delta=0\), corresponding to the coalescence of turning points and the tangency of the energy level with the effective potential.

In Figure \ref{fig:windows2}, the results demonstrate that the combined effect of geometric curvature, encoded through the twist parameter $w$, and magnetic coupling, governed by $\mathcal{B}$, produces a nontrivial restructuring of the effective potential landscape. The factor $\chi(u)=1+w^2u^2$ introduces a position-dependent inertial renormalization, while the projected gauge potential \eqref{vec-pot} induces a nonlinear momentum shift that becomes increasingly significant at intermediate and large transverse displacements. The resulting effective potential $\mathcal{V}(u)$ exhibits a competition-driven structure in which the number of classically allowed regions (confinement windows) depends sensitively on the relative magnitudes of $\mathcal{B}$, $w$, and $\mathcal{E}$. In weak magnetic or strong geometric regimes, multiple disconnected confinement windows may emerge, indicating coexistence of dynamically separated classical trajectories. In contrast, for sufficiently strong magnetic coupling, the potential develops a single dominant well, leading to a unique confinement region. The stationary points defined by $\frac{d\mathcal{V}}{du}=0$ act as transition markers between distinct dynamical regimes and correspond to the onset of qualitative changes in phase-space topology. These points organize the structure of the energy manifold and determine the merging or splitting of confinement regions. The system exhibits a curvature--gauge induced confinement mechanism in which geometry and magnetic effects jointly control the global phase-space accessibility structure.

In the asymptotic regime $|u|\to\infty$, moreover, the metric factor behaves as $\chi(u)\sim w^2u^2$, and the effective potential admits the expansion
\begin{equation}
\mathcal{V}(u)\sim
\frac{q^2\mathcal{B}^2}{8\mu}u^2
-\frac{q\mathcal{B}\mathcal{P}_v}{2\mu w}
-\frac{q^2\mathcal{B}^2}{8\mu w^2}
+\frac{\mathcal{P}_v^2}{2\mu w^2u^2}.\label{pot-expansion}
\end{equation}
The leading quadratic growth demonstrates that large transverse excursions are dynamically suppressed by an emergent confinement mechanism originating from the combined effect of curvature-induced metric modulation and gauge-induced momentum deformation. The constant and inverse-square corrections constitute subleading contributions that shift the effective energy scale without modifying the dominant quadratic confinement behavior. The global dynamics is therefore governed by the competition between metric growth $w^2u^2$ and gauge-induced quadratic momentum deformation. Their relative scaling determines the number of real solutions of the equation $\mathcal{E}=\mathcal{V}(u)$, thereby classifying the dynamics into confined, critical, or transport-dominated regimes. As shown in Figure~\ref{fig:phase-space}, the energy-resolved phase-space portraits indicate that the helicoidal constraint reorganizes Hamiltonian level sets through the combined effects of metric-induced inertial modulation and gauge-induced momentum deformation. The factor $\chi(u)=1+w^2u^2$ introduces a position-dependent effective mass that suppresses large-amplitude transverse motion, while the quadratic gauge potential \eqref{vec-pot} produces a nonlinear shift in the conserved canonical momentum $P_v$, resulting in asymmetric deformation of phase-space trajectories. The resulting dynamics exhibits a continuous foliation of the $(u,p_u)$ phase space into energy-dependent invariant manifolds that remain preserved under symplectic evolution. In the weak-coupling regime, phase-space structures are broad and weakly distorted, reflecting geometry-dominated dynamics in which inertial response comes from the metric tensor. As either the magnetic field $\mathcal{B}$ or the twist parameter $w$ increases, these invariant manifolds undergo progressive compression and anisotropic bending, indicating strengthened coupling between longitudinal gauge momentum and transverse confinement. In the strong-coupling regime, the phase space develops sharply curved and highly structured energy shells, corresponding to effective confinement channels in which geometric and magnetic effects reinforce each other. The persistence of smooth energy gradients across all regimes confirms the Hamiltonian consistency of the numerical scheme and demonstrates that the observed phase-space deformation is non-dissipative, arising purely from deterministic curvature--gauge dynamics on a non-Euclidean configuration space.

\section{Helicoidal Dynamics, Stability, and Quantization} \label{quantization}

The constrained dynamics of a charged particle on a helicoidal manifold admits an exact reduction to a single effective degree of freedom associated with the transverse coordinate $u$, while the longitudinal cyclic variable is eliminated through conservation of its conjugate momentum $\mathcal{P}_v$. This reduction is exact and does not rely on perturbative assumptions, since the symmetry of the embedding guarantees an invariant submanifold in phase space. The resulting system is a nonlinear Hamiltonian flow in one dimension, where all geometric and electromagnetic effects are encoded in an effective potential generated jointly by the induced metric and the projected gauge field. The longitudinal metric component is $g_{vv}=1+w^2u^2$, while $g_{uu}=1$, implying that geometry enters through the metric-induced inertial factor in the longitudinal sector. The structure of $\mathcal{V}(u)$ shows that curvature and magnetic field do not contribute additively but enter through a nonlinear coupling between a metric-induced inertial enhancement and a gauge-induced momentum shift. The dynamics is constrained to the energy shell, which yields the exact relation in Eq. \eqref{eq:pu}, implying that real motion exists only in regions where $\mathcal{E}\ge \mathcal{V}(u)$. This inequality defines a self-consistent confinement domain whose boundaries are determined dynamically rather than imposed externally. The sign structure reflects time-reversal symmetry of the Hamiltonian flow. Hamilton’s equation $\dot{u}=p_u/\mu$ leads directly to the temporal parametrization \cite{guvendi2026charged,goldstein1950classical}
\begin{equation}
dt=\frac{\mu}{p_u}\,du=\sqrt{\frac{\mu}{2}}\frac{du}{\sqrt{\mathcal{E}-\mathcal{V}(u)}}
\end{equation}
which shows that the helicoidal geometry does not modify the kinetic measure but reshapes the effective potential landscape governing temporal evolution. Turning points are defined by $\mathcal{E}=\mathcal{V}(u_\pm)$ and correspond to vanishing kinetic energy, marking the reversal of motion in configuration space. The oscillation period can be obtained from a full cycle between turning points \cite{goldstein1950classical},
\begin{equation}
T(\mathcal{E})=\sqrt{2\mu}\int_{u_-}^{u_+}\frac{du}{\sqrt{\mathcal{E}-\mathcal{V}(u)}}
\end{equation}
which emphasizes that all nonlinear effects of curvature and gauge coupling are encoded exclusively in $\mathcal{V}(u)$. Near a separatrix energy $\mathcal{E}_c$, where a stable and unstable equilibrium coalesce, the potential admits the local expansion
\begin{equation}
\mathcal{V}(u)\simeq \mathcal{E}_c+\frac{1}{2}\mathcal{V}''(u_s)(u-u_s)^2
\end{equation}
leading to a universal logarithmic divergence of the period
\begin{equation}
T(\mathcal{E})\sim \frac{1}{\sqrt{|\mathcal{V}''(u_s)|}}\ln|\mathcal{E}-\mathcal{E}_c|^{-1}
\end{equation}
which reflects critical slowing down governed solely by the local curvature of the effective energy landscape. The invariant characterization of motion is provided by the action variable \cite{goldstein1950classical,pollak2026improved}
\begin{equation}
J(\mathcal{E})=\oint p_u\,du=2\int_{u_-}^{u_+}\sqrt{2\mu(\mathcal{E}-\mathcal{V}(u))}\,du
\end{equation}
which remains conserved under canonical transformations and fully encodes the phase-space geometry of the orbit. Its derivative yields the period
\begin{equation}
\frac{dJ}{d\mathcal{E}}=T(\mathcal{E})
\end{equation}
establishing the action as the generator of the nonlinear frequency map $\Omega(\mathcal{E})=d\mathcal{E}/dJ$. Transport along the longitudinal direction arises from averaging the projected velocity over one oscillation cycle,
\begin{equation}
\langle \dot{v} \rangle=\frac{1}{T(\mathcal{E})}\int_0^{T(\mathcal{E})}\dot{v}(t)\,dt
\end{equation}
which, after transformation to configuration space, becomes
\begin{equation}
\langle \dot{v} \rangle=\frac{1}{T(\mathcal{E})}\int_{u_-}^{u_+}\frac{\mathcal{P}_v-\alpha u^2}{(1+w^2u^2)\sqrt{2\mu(\mathcal{E}-\mathcal{V}(u))}}\,du,\quad \alpha\equiv \frac{q\mathcal{B}w}{2},
\end{equation}
showing explicitly that transport is controlled by competition between magnetic momentum shifting and geometric inertia enhancement. The factor $(1+w^2u^2)$ suppresses longitudinal propagation at large transverse displacement, while the magnetic term introduces a position-dependent drift mechanism. The resulting dynamics exhibits regimes of enhanced transport, trapping, or suppression depending on the relative magnitude of curvature and magnetic coupling. The global phase-space structure is classified by the topology of $\mathcal{V}(u)$: bounded periodic motion for $\mathcal{E}<\mathcal{E}_c$, separatrix dynamics with divergent period at $\mathcal{E}=\mathcal{E}_c$, and unbounded drift for $\mathcal{E}>\mathcal{E}_c$. In the semiclassical regime, quantization follows from the Bohr--Sommerfeld condition \cite{goldstein1950classical,pollak2026improved}
\begin{equation}
J(\mathcal{E}_n)=2\pi\hbar\left(n+\frac{1}{2}\right),\qquad n\in\mathbb{N}_0
\end{equation}
which discretizes allowed energies according to phase-space volume. Near a stable equilibrium point defined by $\mathcal{V}'(u_\ast)=0$ and $\mathcal{V}''(u_\ast)>0$, the potential reduces to
\begin{equation}
\mathcal{V}(u)\simeq \mathcal{V}(u_\ast)+\frac{1}{2}\mathcal{V}''(u_\ast)(u-u_\ast)^2
\end{equation}
yielding harmonic motion with frequency
\begin{equation}
\omega_\ast=\sqrt{\frac{\mathcal{V}''(u_\ast)}{\mu}}
\end{equation}
and locally linear action
\begin{equation}
J(\mathcal{E})\simeq \frac{2\pi}{\omega_\ast}(\mathcal{E}-\mathcal{V}(u_\ast))
\end{equation}
which produces the discrete spectrum
\begin{equation}
\mathcal{E}_n=\mathcal{V}(u_\ast)+\hbar\omega_\ast\left(n+\frac{1}{2}\right)
\end{equation}
with level spacing
\begin{equation}
\Delta\mathcal{E}=\hbar\omega_\ast
\end{equation}
demonstrating that quantum structure inherits its full dependence on geometry and gauge coupling through $\mathcal{V}''(u_\ast)$. Finally, the oscillation frequency is determined directly from the full nonlinear potential curvature,
\begin{equation}
\omega_\ast=\sqrt{\frac{1}{\mu}\left.\frac{d^2}{du^2}\left[\frac{(\mathcal{P}_v-\alpha u^2)^2}{2\mu(1+w^2u^2)}\right]\right|_{u=u_\ast}}
\end{equation}
which makes explicit that spectral and transport properties are governed by inseparable coupling between helicoidal geometry and electromagnetic interaction, producing a unified nonlinear phase-space structure rather than independent geometric or magnetic contributions.

\subsection{Semiclassical Schrödinger Reduction and Helicoidal Asymptotic Landau Quantization with Magnetic-Length Renormalization}\label{sch-mapping}

We consider a particle of mass $\mu$ and charge $q$ constrained to a helicoidal surface with induced metric \(g_{ab}=\mathrm{diag}(1,1+w^2u^2)\), subjected to a uniform magnetic field $\mathcal{B}$ oriented along the embedding axis. The coordinate $v$ is cyclic, and its conjugate momentum $\mathcal{P}_v=\hbar k_v$ is conserved, reducing the full two-dimensional dynamics to an effective one-dimensional spectral problem in the transverse coordinate $u$. In this reduction, the combined effects of curvature and minimal coupling are encoded in the effective potential $\mathcal{V}(u)$ defined in Eq.~\eqref{eff-pot}. The corresponding semiclassical dynamics is governed by the effective Schrödinger Hamiltonian
\begin{equation}
\hat{\mathcal{H}}=-\frac{\hbar^2}{2\mu}\frac{d^2}{du^2}+\mathcal{V}(u).
\label{QH}
\end{equation}
This Hamiltonian represents the leading-order semiclassical quantization of the constrained dynamics, obtained after reduction to the invariant coordinate $u$. The potential structure reflects a competition between magnetic confinement and geometric dilation: the gauge-covariant momentum shift enters through the numerator, while the helicoidal metric modifies the effective kinetic term via the $(1+w^2u^2)^{-1}$ factor. In the asymptotic regime $|u|\to\infty$, the metric factor admits the expansion
\begin{equation}
(1+w^2u^2)^{-1}=(w^2u^2)^{-1}\left[1-\frac{1}{w^2u^2}+\mathcal{O}(u^{-4})\right],
\end{equation}
while the squared covariant momentum behaves as
\begin{equation}
\left(\hbar k_v-\frac{q\mathcal{B}}{2}wu^2\right)^2=
\left(\frac{q\mathcal{B}}{2}w\right)^2u^4
-q\mathcal{B}w\,\hbar k_v\,u^2
+\hbar^2k_v^2.
\end{equation}
Consistently combining these expansions yields a hierarchy of contributions in descending powers of $u$, leading to an effective potential of the form
\begin{equation}
\mathcal{V}(u)=
\frac{q^2\mathcal{B}^2}{8\mu}u^2
-\frac{q\mathcal{B}\hbar k_v}{2\mu w}
+\frac{\hbar^2k_v^2}{2\mu w^2u^2}
-\frac{q^2\mathcal{B}^2}{8\mu w^2}
+\mathcal{O}(u^{-4}).
\end{equation}
The leading term is purely harmonic and independent of the helicoidal parameter $w$, whereas the helicoidal geometry contributes through constant spectral shifts and inverse-square corrections. Introducing the cyclotron frequency and magnetic length,
\begin{equation}
\omega_c=\frac{|q\mathcal{B}|}{\mu},
\qquad
\ell_\mathcal{B}=\sqrt{\frac{\hbar}{|q\mathcal{B}|}},
\end{equation}
the dominant contribution can be rewritten as
\begin{equation}
\mathcal{V}_{\mathrm{lead}}(u)=
\frac{1}{2}\mu\left(\frac{\omega_c}{2}\right)^2u^2.
\end{equation}
This identifies the effective asymptotic frequency
\begin{equation}
\omega_{\mathrm{eff}}=\frac{\omega_c}{2}.
\end{equation}
The corresponding localization length is therefore
\begin{equation}
\ell=\sqrt{\frac{\hbar}{\mu\omega_{\mathrm{eff}}}}
=\sqrt{2}\,\ell_\mathcal{B},
\end{equation}
showing that the helicoidal geometry renormalizes the magnetic length without introducing an additional intrinsic scale at leading order. In the asymptotic region, the Hamiltonian in Eq.~\eqref{QH} admits the decomposition
\begin{equation}
\hat{\mathcal{H}}=
\left[
-\frac{\hbar^2}{2\mu}\frac{d^2}{du^2}
+\frac{1}{2}\mu\omega_{\mathrm{eff}}^2u^2
-\frac{q\mathcal{B}\hbar k_v}{2\mu w}
-\frac{q^2\mathcal{B}^2}{8\mu w^2}
\right]
+\frac{\hbar^2k_v^2}{2\mu w^2u^2}
+\mathcal{O}(u^{-4}),
\end{equation}
where the first bracket defines the leading asymptotic Hamiltonian $\hat{\mathcal{H}}_{\mathrm{asy}}$, while the remaining terms constitute subleading semiclassical corrections that vanish asymptotically. Since $u^{-2}\ll u^2$ in this regime, the inverse-square contribution affects only higher-order spectral corrections and does not modify the leading harmonic oscillator structure, apart from subleading energy shifts. Accordingly, the asymptotic eigenvalue problem reduces to
\begin{equation}
\left[
-\frac{\hbar^2}{2\mu}\frac{d^2}{du^2}
+\frac{1}{2}\mu\omega_{\mathrm{eff}}^2u^2
\right]\psi
=
\left(
\tilde{\mathcal{E}}
+\frac{q\mathcal{B}\hbar k_v}{2\mu w}
+\frac{q^2\mathcal{B}^2}{8\mu w^2}
\right)\psi.\label{AEV}
\end{equation}
The spectrum is therefore that of a shifted quantum oscillator,
\begin{equation}
\tilde{\mathcal{E}}_n
=
\hbar\omega_{\mathrm{eff}}
\left(n+\frac{1}{2}\right)
-\frac{q\mathcal{B}\hbar k_v}{2\mu w}
-\frac{q^2\mathcal{B}^2}{8\mu w^2},
\qquad
n\in\mathbb{N}_0.\label{SQO}
\end{equation}
Thus, the helicoidal geometry contributes a nontrivial curvature-induced spectral shift that is required for consistency with Eqs.~\eqref{pot-expansion}, \eqref{AEV}, and \eqref{SQO}. Within this leading-order semiclassical Schrödinger reduction, the asymptotic dynamics reduces to a Landau-type harmonic quantization with effective frequency $\omega_{\mathrm{eff}}=\omega_c/2$, where the helicoidal geometry manifests through a renormalized magnetic length, constant spectral shifts, and subleading inverse-square corrections that vanish in the limit $|u|\to\infty$.

\subsection{Geometry-induced phase transition}\label{subsec:phase_transition}

\begin{figure*}[!t]
\centering
\includegraphics[scale=0.50]{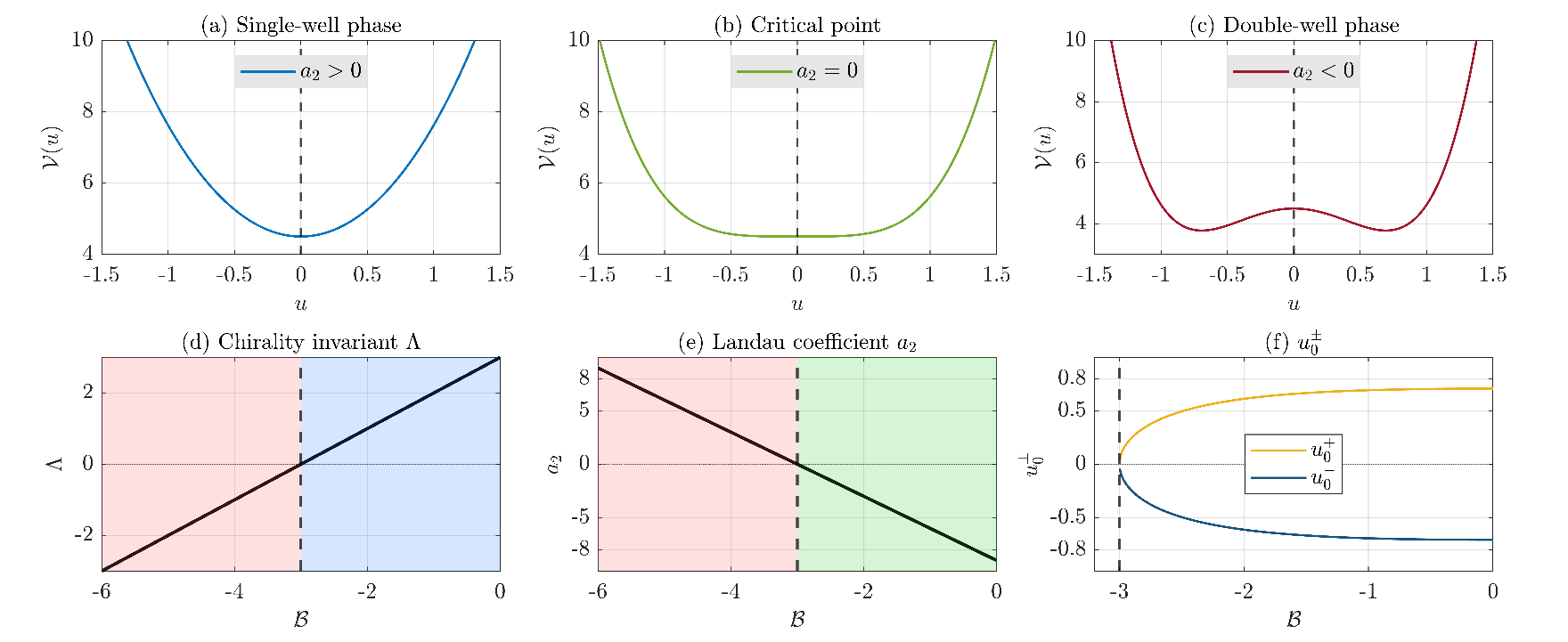}
\caption{\fontsize{7.6}{8.6}\selectfont Geometry-induced quantum phase transition and chirality inversion in a helicoidal quantum system. All results are obtained with $\hbar=1$, $\mu=1$, $q=1$, and $w=1$, for a helicoidal momentum $k_v=3$. The critical magnetic field is $\mathcal{B}_c = -\hbar k_v w/q = -3$, and all results are evaluated in the interval $\mathcal{B} \in [\mathcal{B}_c - 3,\, \mathcal{B}_c + 3]$ centered at the transition point. The effective potential is given by $\mathcal{V}(u)=\frac{1}{2\mu}(a_0+a_2 u^2+a_4 u^4)$, where $a_2=-\hbar k_v w\,\Lambda$ and $\Lambda=q\mathcal{B}+\hbar k_v w$. Panels (a)--(c) show the evolution of the effective potential across the quantum phase transition: (a) the single-well regime ($a_2>0$), (b) the critical point ($a_2=0$), and (c) the double-well regime ($a_2<0$), where spontaneous symmetry breaking $u\rightarrow -u$ occurs. In the broken-symmetry phase, two degenerate minima emerge at $u_0^\pm=\pm\sqrt{|a_2|/(2a_4)}$, corresponding to opposite chirality sectors. Panel (d) shows the chirality invariant $\Lambda$, which crosses zero linearly at $\mathcal{B}_c$, separating two distinct chirality regimes governed by magnetic--geometric competition. Panel (e) shows the Landau coefficient $a_2$, which changes sign at the same critical field, marking a second-order quantum phase transition. Panel (f) shows the order parameter bifurcation structure $u_0^\pm$, explicitly demonstrating the emergence of degenerate ground states in the symmetry-broken phase. The quartic coefficient $a_4>0$ remains strictly positive for all $\mathcal{B}$, ensuring global stability of the effective Hamiltonian.}
\label{fig:QPT}
\end{figure*}
We derive the quantum phase structure directly from the reduced Hamiltonian in Eq.~\eqref{QH}, with effective potential given in Eq.~\eqref{eff-pot}. The critical behavior is obtained from a controlled small-$u$ expansion about the symmetry point $u=0$, invariant under $u\rightarrow -u$. Setting $\mathcal{P}_v=\hbar k_v$, the gauge sector expands exactly as
\begin{equation}
\left(\hbar k_v-\frac{q\mathcal{B}}{2}wu^2\right)^2
= \hbar^2 k_v^2
- q\mathcal{B}\hbar k_v w\,u^2
+\frac{q^2\mathcal{B}^2}{4}w^2u^4,
\end{equation}
while the dimensionless geometric factor admits the convergent expansion for $|wu|\ll 1$,
\begin{equation}
\frac{1}{1+w^2u^2}
=1-w^2u^2+w^4u^4+\mathcal{O}(u^6).
\end{equation}
Substituting into Eq.~\eqref{eff-pot} and collecting terms up to $\mathcal{O}(u^4)$ yields the Landau-type expansion
\begin{equation}
\mathcal{V}(u)=\frac{1}{2\mu}\left(a_0+a_2 u^2+a_4 u^4\right)+\mathcal{O}(u^6),
\end{equation}
with coefficients
\begin{align}
a_0 &= \hbar^2 k_v^2, \qquad
a_2 = -\hbar k_v w\left(q\mathcal{B}+\hbar k_v w\right),\\
a_4 &= w^2\left[\frac{q^2\mathcal{B}^2}{4} + q\mathcal{B}\hbar k_v w + \hbar^2 k_v^2 w^2\right].
\end{align}
The quadratic coefficient $a_2$ factorizes exactly as
\begin{equation}
a_2 = -\hbar k_v w\,\Lambda,
\qquad
\Lambda \equiv q\mathcal{B}+\hbar k_v w,
\end{equation}
identifying $\Lambda$ as the curvature--gauge interference invariant controlling the local stability of the symmetric configuration $u=0$. The quartic coefficient satisfies $a_4>0$ for all real parameter values, ensuring boundedness of the effective potential within the validity of the expansion. The effective Hamiltonian exhibits an emergent chirality structure originating from the signed competition between magnetic and geometric contributions (see also \cite{saxena-1,saxena-2}). This structure is encoded in the invariant $\Lambda$, which combines a charge-dependent term $q\mathcal{B}$ with a geometry-induced momentum--connection coupling $\hbar k_v w$. The sign of $\Lambda$ defines an effective chirality sector, while the factor $\hbar k_v w$ encodes the intrinsic handedness of motion along the helicoidal embedding. Under reversal of propagation direction $k_v \rightarrow -k_v$ or charge conjugation $q\rightarrow -q$, the invariant transforms as
\begin{equation}
\Lambda \rightarrow q\mathcal{B} - \hbar k_v w \quad \text{or} \quad \Lambda \rightarrow -q\mathcal{B} + \hbar k_v w,
\end{equation}
demonstrating that the system admits a tunable chirality imbalance controlled jointly by charge and geometry. The sign structure of $a_2$ therefore determines not only stability but also the effective chirality sector of the ground state. The harmonic sector is determined by the local curvature of the effective potential. In the quadratic regime the effective frequency is defined as
\begin{equation}
\omega^2=\frac{a_2}{\mu^2}, \qquad [a_2/\mu^2]=T^{-2},
\end{equation}
For $a_2>0$, the system realizes a single-well confining phase (see Figure \ref{fig:QPT}) with harmonic excitation spectrum
\begin{equation}
\mathcal{E}_n=\frac{\hbar}{\mu}\sqrt{a_2}\left(n+\frac{1}{2}\right),
\end{equation}
where $\frac{\hbar}{\mu}\sqrt{a_2}$ has dimensions of energy. For $a_2<0$, the symmetric point becomes unstable and quartic stabilization generates a double-well structure with minima (see Figure \ref{fig:QPT})
\begin{equation}
u_0^2=\frac{|a_2|}{2a_4},
\end{equation}
corresponding to spontaneous breaking of the discrete symmetry $u\rightarrow -u$. In this regime, the system selects one of two symmetry-related minima, which correspond to opposite effective chirality sectors of the reduced dynamics. The quantum critical point is defined by
\begin{equation}
a_2=0 \quad \Longleftrightarrow \quad \Lambda=0,
\qquad
q\mathcal{B}+\hbar k_v w=0,
\end{equation}
which yields the critical magnetic field
\begin{equation}
\mathcal{B}_c = -\frac{\hbar k_v w}{q}.
\end{equation}
At this point, and within $|wu|\ll 1$, the Hamiltonian reduces to the scale-invariant quartic theory
\begin{equation}
\hat{\mathcal{H}}_c=
-\frac{\hbar^2}{2\mu}\frac{d^2}{du^2}
+\frac{a_4}{2\mu}u^4.
\end{equation}
Ground-state fluctuations in the harmonic phase follow from the curvature scale \cite{Gottfried},
\begin{equation}
\langle u^2\rangle=\frac{\hbar}{2\sqrt{a_2}},
\qquad a_2>0,
\end{equation}
where the relation $\omega=\sqrt{a_2}/\mu$ (equivalently $\omega^2=a_2/\mu^2$) ensures consistency of the effective harmonic reduction. This implies the critical scaling \cite{Gottfried}
\begin{equation}
\langle u^2\rangle \sim |a_2|^{-1/2} \sim |\Lambda|^{-1/2}.
\end{equation}
This establishes a second-order quantum phase transition governed entirely by the invariant $\Lambda$, defining a codimension-one critical hypersurface in the parameter space $(\mathcal{B},k_v,w)$. The transition originates from the exact competition between the magnetic contribution $q\mathcal{B}$ and the geometry-induced momentum scale $\hbar k_v w$, with the helicoidal embedding actively controlling both the stability and the emergent chirality structure of the quantum phase through $\Lambda$.

In Figure \ref{fig:QPT}, the system exhibits a geometry-driven second-order quantum phase transition governed by the chirality--geometry invariant $\Lambda=q\mathcal{B}+\hbar k_v w$. The critical magnetic field is
\begin{equation}
\mathcal{B}_c=-\frac{\hbar k_v w}{q}=-3,
\end{equation}
which defines a codimension-one transition point separating two distinct chirality sectors distinguished by the sign of $\Lambda$. For $\mathcal{B}>\mathcal{B}_c$, the invariant satisfies $\Lambda>0$, implying $a_2<0$ due to $a_2=-\hbar k_v w\,\Lambda$. In this regime, the effective potential is in a single-well configuration with a stable symmetric ground state at $u=0$. The system preserves the discrete symmetry $u\rightarrow -u$, and fluctuations are governed by a locally harmonic curvature around the origin. For $\mathcal{B}<\mathcal{B}_c$, the condition $\Lambda<0$ yields $a_2>0$, and the symmetric configuration becomes dynamically unstable. The effective potential undergoes a Landau-type bifurcation into a double-well structure with two degenerate minima at
\begin{equation}
u_0^\pm=\pm\sqrt{\frac{|a_2|}{2a_4}},
\end{equation}
signaling spontaneous breaking of the discrete symmetry $u\rightarrow -u$. Although the quadratic sector becomes unstable in this regime, the quartic coefficient $a_4>0$ guarantees boundedness of the Hamiltonian and stabilizes the broken-symmetry phase. The chirality inversion is controlled by the competition between the magnetic contribution $q\mathcal{B}$ and the intrinsic geometric coupling $\hbar k_v w$, encoded in the invariant $\Lambda$. This competition reverses the sign of $\Lambda$ across $\mathcal{B}_c$ and induces a controlled chirality transition, where the system selects one of the degenerate minima $u_0^\pm$, leading to spontaneous chirality selection in the reduced dynamics.

\section{Conclusion}\label{conc}

\setlength{\parindent}{0pt}

We have developed a geometric--dynamical framework for charged particle motion on a helicoidal Riemannian manifold embedded in $\mathbb{R}^3$, in the presence of a uniform magnetic field defined in the ambient Euclidean space. The construction is based on an exact embedding map, from which both the induced metric and the intrinsic gauge structure are derived by pullback of the ambient Maxwell field. This guarantees that all electromagnetic effects on the surface are geometrically consistent and free from coordinate artifacts. The resulting configuration space is a non-Euclidean manifold with metric $ds^2 = du^2 + (1+w^2u^2)dv^2$, where the twist parameter $w=2\pi m/L$ provides a direct geometric control knob linking global topology (number of helical windings) to local dynamical anisotropy.

\vspace{0.01cm}

A central outcome of this work is the exact reduction of the full constrained dynamics to an effective one-dimensional Hamiltonian system in the transverse coordinate $u$, governed by a nonlinear effective potential $\mathcal{V}(u)$ encoding inseparable curvature--gauge coupling. As shown in Section~\ref{CHF} and Figure~\ref{fig:-1}, the helicoidal geometry induces a position-dependent inertial renormalization through $\chi(u)=1+w^2u^2$, while the projected magnetic vector potential generates a nonlinear momentum shift proportional to $u^2$. Their combined action  produces an effective potential landscape whose qualitative structure is controlled by two competing scales: the geometric deformation scale $w$ and the magnetic scale $\mathcal{B}$.

\vspace{0.01cm}

The resulting dynamics exhibits a rich hierarchy of regimes. In the weak-coupling limit, motion is predominantly geometry-dominated, characterized by broad phase-space structures and multiple confinement windows (Figure~\ref{fig:windows2}). In contrast, strong magnetic coupling suppresses transverse excursions and collapses the accessible configuration space into a single dominant confinement region. The exact turning-point analysis in Section~\ref{PS} reveals that the topology of phase-space trajectories is governed by a quadratic constraint in $u^2$, whose discriminant structure determines whether the system supports one or multiple dynamically disconnected invariant regions. This establishes a precise correspondence between algebraic properties of $\mathcal{V}(u)$ and global phase-space connectivity.

\vspace{0.01cm}

A further key result is the identification of a curvature--gauge controlled asymptotic confinement mechanism. As $|u|\to\infty$, the effective potential becomes universally harmonic with curvature proportional to $\mathcal{B}^2$, while geometry enters through a renormalization of subleading scales. This leads to an emergent magnetic-length rescaling $\ell=\sqrt{2}\,\ell_{\mathcal{B}}$, demonstrating that the helicoidal embedding does not introduce an independent infrared scale but instead renormalizes the magnetic confinement structure. This is reflected in the asymptotic Landau-type spectrum derived in Section~\ref{sch-mapping}, where the effective cyclotron frequency is reduced to $\omega_{\mathrm{eff}}=\omega_c/2$.

\vspace{0.01cm}

At the quantum level, the system exhibits a geometry-driven second-order quantum phase transition controlled by the chirality--geometry invariant
\begin{equation}
\Lambda = q\mathcal{B} + \hbar k_v w,
\end{equation}
which encapsulates the competition between magnetic forcing and helicoidal momentum--connection coupling. As demonstrated in Figure~\ref{fig:QPT}, the sign change of $\Lambda$ induces a Landau-type bifurcation of the effective potential from a single-well to a double-well structure, accompanied by spontaneous symmetry breaking $u\rightarrow -u$ and the emergence of degenerate chirality sectors. The critical point $\mathcal{B}_c=-\hbar k_v w/q$ defines a codimension-one separatrix in parameter space, separating symmetric and symmetry-broken phases. Near this transition, fluctuations diverge with universal scaling $\langle u^2\rangle \sim |\Lambda|^{-1/2}$, confirming the second-order nature of the transition.

\vspace{0.01cm}

From a dynamical perspective, the helicoidal constraint acts as a geometric amplifier of gauge-induced effects: curvature modifies inertial response via the metric tensor, while the magnetic field reshapes momentum space through nonlinear minimal coupling. Their combined action generates a nontrivial reorganization of phase-space foliations, as illustrated in Figure~\ref{fig:phase-space}, where invariant energy shells are continuously deformed but preserved under symplectic flow. The resulting dynamics is fully deterministic, non-dissipative, and structurally stable, yet capable of producing highly intricate confinement and transport phenomena.

\vspace{0.01cm}

Accordingly, the helicoidal manifold emerges as a minimal but nontrivial geometric platform in which curvature and electromagnetism are not separable inputs but mutually coupled dynamical agents. The framework developed here demonstrates that geometric embedding alone can induce effective gauge nonlinearities, renormalize fundamental magnetic scales, and generate quantum critical behavior without the introduction of external potentials beyond a uniform field in the ambient space.

\vspace{0.01cm}

The present construction opens several directions for further analytical and physical development. A first natural extension concerns the inclusion of intrinsic curvature corrections beyond the induced metric approximation, particularly within the thin-layer quantization framework. In this regime, geometry-induced scalar potentials of da Costa type may compete with the gauge-induced contributions identified here, potentially leading to additional confinement channels or curvature-driven topological transitions.

\vspace{0.01cm}

A second direction involves the extension to non-Abelian gauge structures on helicoidal manifolds. Replacing the Abelian magnetic field with more general gauge connections would promote the momentum shift mechanism $\mathcal{A}_v(u)$ into a matrix-valued geometric coupling, enabling the study of curvature-controlled isospin transport and non-Abelian Landau quantization on embedded manifolds. This would naturally generalize the invariant $\Lambda$ into a noncommutative chirality tensor governing internal symmetry breaking.

\vspace{0.01cm}

Third, the semiclassical mapping employed throughout this work suggests a deeper connection to integrable and near-integrable systems. In particular, the effective one-dimensional Hamiltonian in $u$ provides a natural setting for studying chaos--integrability transitions induced by competing curvature and gauge scales. Lyapunov analysis and action-angle constructions could clarify whether helicoidal confinement supports hidden integrable limits or mixed-phase dynamical structures.

\vspace{0.01cm}

Fourth, the quantum phase transition identified via the invariant $\Lambda$ invites a renormalization-group treatment. The emergence of a codimension-one critical manifold suggests that helicoidal geometry may define a universality class of curvature--gauge driven transitions, with scaling exponents controlled by combined geometric and electromagnetic couplings. A systematic renormalization-group flow in $(\mathcal{B},k_v,w)$ space could reveal fixed points associated with symmetric, broken-chirality, and Landau-dominated regimes.

\vspace{0.01cm}

Finally, from an applied perspective, the results suggest potential realizations in engineered curved nanostructures and twisted quantum materials, where effective helicoidal geometries can be synthesized through strain, torsion, or moiré engineering. In such systems, the predicted magnetic-length renormalization, confinement window structure, and chirality inversion transition could be directly probed via transport spectroscopy or quantum interference measurements.

\section*{Acknowledgments}
H.H. is grateful to Excellence project FoS UHK 2203/2025-2026 for the financial support.

\section*{ Credit authorship contribution statement}
\textbf{Abdullah Guvendi}: Conceptualization, Methodology, Formal Analysis, Writing - Original Draft, Investigation, Visualization, Writing - Review and Editing.\\
\textbf{Hassan Hassanabadi}: Conceptualization, Methodology, Formal Analysis, Writing - Original Draft, Investigation, Visualization, Writing - Review and Editing.\\
\textbf{Semra Gurtas Dogan}: Conceptualization, Methodology, Formal Analysis, Writing - Original Draft, Investigation, Visualization, Writing - Review and Editing.\\
\textbf{Omar Mustafa}: Conceptualization, Methodology, Formal Analysis,  Writing - Original Draft, Investigation, Visualization, Writing - Review and Editing.

\section*{Data availability}

This manuscript has no associated data.

\section*{Conflicts of interest statement}

No conflict of interest declared by the authors.



\end{document}